\documentclass[aps,prb,amsmath,amssymb,floatfix,twocolumn]{revtex4}

\usepackage{multirow}
\usepackage{bbold}
\usepackage{graphicx}
\usepackage{subfigure}
\usepackage{color}
\usepackage{hyperref}
\usepackage[justification=raggedright]{caption}

\usepackage[normalem]{ulem}

\usepackage{physics}

\usepackage{mathtools}

\usepackage{amstext}
\usepackage{array}

\usepackage{diagbox}

\usepackage{siunitx}

\usepackage{yfonts}

\allowdisplaybreaks

\usepackage{setspace}

\newcommand{\ncmd}{\newcommand}
\ncmd{\nn}{\nonumber}
\ncmd{\pg}[1]{\textcolor{red}{#1}}
\ncmd{\mbf}[1]{\bs{#1}}
\ncmd{\Lam}{\Lambda}
\ncmd{\lam}{\lambda}
\ncmd{\Gam}{\Gamma}
\ncmd{\gam}{\gamma}
\ncmd{\sig}{\sigma}
\ncmd{\Dl}{\Delta}
\ncmd{\dl}{\delta}
\ncmd{\kap}{\kappa}
\ncmd{\mc}{\mathcal}
\ncmd{\veps}{\varepsilon}
\ncmd{\vphi}{\varphi}
\ncmd{\vtheta}{\vartheta}
\ncmd{\note}[1]{{\color{red}{\bf{#1} } } }
\ncmd{\new}[1]{{\color{blue}{\texttt{#1}  } } }
\ncmd{\eq}[1]{Eq. \eqref{#1}}
\ncmd{\bs}{\boldsymbol}
\ncmd{\pll}{\parallel}

\begin{document}

\title{Mechanism of skyrmion condensation and pairing for twisted bi-layer graphene}
\author{Dian Jing$^{1,2,3}$, Alexander Conkey Tyner$^4$, Pallab Goswami$^{1,4}$}
\affiliation{$^{1}$ Department of Physics and Astronomy, Northwestern University, Evanston, IL 60208}
\affiliation{$^{2}$ Department of Chemical and Biological Engineering, Northwestern University, Evanston, IL 60208}
\affiliation{$^{3}$ Integrated Science Program, Northwestern University, Evanston, IL 60208}
\affiliation{$^{4}$ Graduate Program in Applied Physics, Northwestern University, Evanston, IL 60208}

\date{\today}

\begin{abstract}
When quantum flavor Hall insulator phases of itinerant fermions are disordered by strong quantum fluctuations, the condensation of skyrmion textures of order parameter fields can lead to superconductivity. In this work, we address the mechanism of skyrmion condensation by considering the scattering between (2+1)-dimensional, Weyl fermions and hedgehog type tunneling configurations of order parameters that violate the skyrmion-number conservation law. We show the quantized, flavor Hall conductivity ($\sigma^f_{xy}$) controls the degeneracy of topologically protected, fermion zero-modes, localized on hedgehogs, and the overlap between zero-mode eigenfunctions or 't Hooft vertex determines the nature of pairing. We demonstrate the quantum-disordered, flavor Hall insulators with $\sigma^f_{xy}= 2 N$ lead to different types of charge $2 N e^-$ superconductivity. Some implications for the competition among flavor Hall insulators, the charge $2e^-$ paired states in BCS and pair-density-wave channels, and the composite, charge $4e^-$ superconductors for twisted bilayer graphene are outlined. 
\end{abstract}

\maketitle

\underline{\emph{Introduction} }: Skyrmions are smooth, topologically non-trivial, textures of three-component, unit vector field $\hat{\bs{\Omega}}(\mathbf{x})$ of $O(3)$ non-linear sigma model (NLSM) at two spatial dimensions.~\cite{Rajaraman} Since $\hat{\bs{\Omega}}(\mathbf{x})$ is equivalent to a two-sphere ($S^2$), the skyrmion configurations are classified by the second homotopy group: $\pi_2(S^2)=\mathbb{Z}$. 
Inside the ordered phase of NLSM, the skyrmion requires a finite creation energy $E_{sk} = 4\pi \rho_s |W_{sk}|$, where $\rho_s$ is the spin-stiffness of the NLSM, and $W_{sk}$ is the integer-valued, winding number. If $\hat{\bs{\Omega}}(\mathbf{x})$ is coupled to (2+1)-dimensional Weyl fermions as a mass type order parameter, the skyrmions can support induced ``fermion numbers", indicating the presence of fluctuating competing orders inside the skyrmion core.~\cite{Wilczek1, Jaroszewicz1, Jaroszewicz2, Wilczek2, Abanov, TanakaHu, FisherSenthil, Grover, YaoLee, FuSachdev, RoyHerbut, Herbut1, Moon, Chakravarty1, GoswamiSi2, GoswamiSi3,GoswamiSi4} When the NLSM is disordered by quantum fluctuations, the vanishing of excitation gap can allow the condensation of skyrmions. Hence, the quantum-disordered, anti-ferromagnet (AFM) and quantum spin Hall (QSH) states can support nucleation of spin-singlet, valence bond solids (VBS)~\cite{FisherSenthil} and $s$-wave superconductivity,~\cite{Grover,Herbut1,Moon,Chakravarty1} respectively.  

The analysis of such competing orders is often guided by the construction of $SO(5)$ NLSMs that treat spin-triplet and spin-singlet orders on an equal footing. After integrating out the gapped fermion fields, one obtains a non-Abelian, Berry's phase for the five-component vector field~\cite{Abanov,TanakaHu, FisherSenthil, Grover, YaoLee, FuSachdev, RoyHerbut, Herbut1, Moon, Chakravarty1, GoswamiSi2}, which is also known as the Wess-Zumino-Witten (WZW) term. The relationship between $SO(5)$ WZW model and the unconventional, continuous quantum phase transitions between two ordered states~\cite{Senthiletal2} is being actively studied by many groups.~\cite{Chalker1,CWang,ZWang} 

Owing to the discovery of superconductivity in twisted bilayer graphene (TBLG) in the vicinity of correlated insulating states, the analysis of competing particle-hole and particle-particle orders of Weyl fermions has become an  important, physically relevant problem.~\cite{Cao1,Taniguchi,Efetov1,Efetov2,Saito,Kerelsky,Choi,Jiang,Xie,Wong,Zondiner,LiuWang,Cao2} The interactions between fermion fields and bosonic collective modes, such as phonons and smooth order parameter fluctuations are being vigorously investigated, as potential candidates for pairing mechanism.~\cite{Wu,Isobe,Lian,You,Classen,Chubukov,Kozil,Fernandes2} Recently, the skyrmion condensation has also been considered for addressing the competition between correlated insulators and adjacent superconducting states.~\cite{Zaletel1,Christos,Zaletel2,Khalaf} Based on the low energy theories of $4-$ and $8-$ flavors of Weyl fermions, different types of level-$1$~\cite{Zaletel1} and level-$2$~\cite{Christos} $SO(5)$ WZW models have been proposed for the integer filling fractions $\nu=2$, and $0$, respectively. The topological pairing mechanism can provide critical, non-perturbative insights into the nature of competing orders and emergent symmetries, which are generally inaccessible through perturbation theories about topologically trivial configurations of collective modes. This motivates us to ask the following question. 

How do the skyrmions condense and determine the nature of competing orders? Due to the vanishing of $E_{sk}$, all topologically distinct ground states, labeled by different $W_{sk}$'s can become energetically degenerate. Such infinite degeneracy of ground states is removed by hedgehog or monopole type tunneling configurations of order parameter in Euclidean space-time. They violate the skyrmion-number conservation law and lead to a superposition of different skyrmion configurations as the true ground state.~\cite{Duncan,MurthySachdev,Read} Therefore, the mechanism of condensation and the nature of superconducting states cannot be clearly understood, without considering interactions between Weyl fermions and hedgehogs. 

Furthermore, for $4N$-flavors of Weyl fermions with $N>1$, a spin-triplet order has many competing spin-singlet, mass orders, which anti-commute with each other. Naturally, there are multiple candidates for level-$N$ $SO(5)$ WZW theories, and they cannot provide an unbiased description of competing spin-singlet orders. Such issues for the $N=2$ case were identified in Refs.~\onlinecite{GoswamiSi2, GoswamiSi3}, while addressing the competition between AFM, VBS, and Kondo-singlets. Ref.~\onlinecite{GoswamiSi3} has showed the space-time dependent, Weyl operator in the presence of hedgehog configurations of AFM order parameter possesses topologically protected fermion zero-modes. Many similarities between this problem and (3+1)-dimensional quantum chromodynamics~\cite{Shuryak,Diakonov} were pointed out. The overlap between zero-modes was shown to cause an effective $2N$-fermion interaction or 't Hooft vertex (TV).~\cite{'T Hooft} For $N=1$, TV is a frequency-momentum dependent fermion billinear, describing VBS order. For $N \geq 2$, the quartic TV corresponds to a composite, spin-singlet order, which cannot be described by VBS or Kondo-singlet type fermion bilinears. 

The primary goal of our current work is to address the nature of paired states arising from the condensation of skyrmion textures of quantum flavor Hall (QFH) insulators for $4N$-flavors of Weyl fermions, by explicitly considering the role of hedgehogs. For the clarity of presentation, we will only consider triplet orders for one of the $SU(2)$ sub-groups of $SU(4N)$. In response to the externally applied electric fields, uniformly ordered, QFH states exhibit quantized Hall conductivity $\sigma^f_{xy}=2N$ for the spin or valley type flavor currents. For TBLG, we will mainly focus on $\nu=0$. Next we set up our notations for the effective theory of Weyl fermions.

\underline{\emph{Flavor symmetry of Weyl fermions}}: The low energy theory of many two-dimensional systems can be described by multiple species of linearly dispersing Weyl fermions. The Weyl points with chirality $+1$ ($-1$) act as unit strength vortices (anti-vortices) in momentum space. In this work, we denote them as the right- and the left- handed Weyl points, respectively, which come in pairs, due to the fermion doubling theorem. On a general ground, we will consider $2N$ pairs of right- and left- handed fermions, where the factor of $2$ ($N$) counts for the spin (layer or other internal) degrees of freedom. We will describe the two-component, right-handed and left-handed fermion fields by $R^T_{i,s}(x_0,\bs x)=(R_{A,i,s,}(x_0,\bs x) , R_{B,i,s}(x_0,\bs x) )$, and $L^T_{i,s}(x_0,\bs x)=(L_{A,i,s}(x_0,\bs x) , L_{B,i,s}(x_0,\bs x) )$, respectively, where $s=\uparrow, \downarrow$ is the spin index, $i=1,...,N$ is the collective index for layer/internal degrees of freedom, and $A/B$ corresponds to the sub-lattice (SL) index. We have absorbed the Fermi velocity $v_F$ on the imaginary time to write $x_0= - i v_F t$. After performing the SL transformation $L_{i,s} \to \tau_2 L_{i,s} $ and combining all degrees of freedom into an $8N$-component spinor $\psi^\dagger_{8N}=(R^\dagger_{1,\uparrow}, R^\dagger_{1,\downarrow},...,R^\dagger_{2N,\uparrow}, R^\dagger_{2N,\downarrow},L^\dagger_{1,\uparrow}, L^\dagger_{1\downarrow},...,L^\dagger_{2N,\uparrow}, L^\dagger_{2N\downarrow}) $, the Euclidean action of $4N$ flavors of Weyl fermions can be written in a manifestly flavor-symmetric form 
\begin{equation}
S_0=\int d^3x \; \bar{\psi}_{8N} \; \mathbb{1}_{4N \times 4N}  \otimes  \left[ \; \sum_{\mu=0}^{2} \; \gamma_{\mu} \; \partial_\mu \right] \; \psi_{8N}, \label{eq1}
\end{equation}
where $\mathbb{1}_{4N \times 4N}$ is the $4N \times 4N$ identity matrix, $\bar{\psi}_{8N}=\psi^\dagger_{8N} \mathbb{1}_{4N \times 4N} \otimes \gamma_0$, $\gamma_0=\tau_3$, $\gamma_1=\tau_2$, $\gamma_2=-\tau_1$. The $2 \times 2$ identity matrix $\tau_0$ and the Pauli matrices $\tau_j$'s with $j=1,2,3$ act on the sub-lattice (SL) index. Notice in the Euclidean frequency-momentum space, the Weyl points appear as the unit strength, hedgehog singularities. \\

The action is invariant under (i) the global $U(1)$ transformation: $\psi_{8N} \to \exp[i \theta \; \mathbb{1}_{4N \times 4N} \otimes \tau_0] \psi_{8N}$,  and (ii) the global, flavor symmetry transformation: $\psi_{8N} \to U \psi_{8N}$, with $U \in SU(4N)$. Hence, the free fermion action possesses $U(1) \times SU(4N)=U(4N)$ symmetry, and we can define the conserved total number current $J^0_{\mu} =\bar{\psi}_{8N} \mathbb{1}_{4N \times 4N} \otimes \gamma_\mu \psi_{8N}$, and the flavor currents 
$
J^i_{\mu}= \bar{\psi}_{8N} \hat{\Lambda}_{i} \otimes \gamma_\mu  \psi_{8N}$, 
where $\hat{\Lambda}_i$ with $i=1,2,..,(4N)^2-1$ are the $4N \times 4N$, Hermitian generators of $SU(4N)$ group. 
For the $N=1$ model of spinful mono-layer graphene (MLG), QFH order can describe the QSH or the quantum valley Hall (QVH) states. The $N=2$ model corresponds to the simplest low energy theory of nearly-flat bands of TBLG.~\cite{Bistritzer,Gail,Santos,Po} For TBLG, the QFH mass can correspond to the QSH, the QVH, or the quantum mini-valley Hall (QMVH) states. Additional Weyl points can also arise from the effects of trigonal warping.~\cite{Geim} The $SU(4N)$ generators can be expressed as suitable linear combinations of direct products of $\log_2(4N)$ sets of identity and Pauli matrices operating on spin and valley type degrees of freedom. While describing MLG and TBLG, we will reserve $\sigma_\lambda$, $\eta_\lambda$, $\rho_\lambda$ respectively for the spin, the valley ($L$ vs. $R$), and the mini-valley ($R_1$ vs $R_2$ within a Moi\'re Brillouin zone) sectors. Next, we consider the skyrmion textures of QFH orders.

\underline{\emph{Skyrmions of QFH orders}}: The $SU(2N)$ symmetry preserving, QSH mass term is given by
$\mathcal{O}_{QSH}= \mathbf{\Omega} (x) \cdot \bar{\psi} \mathbb{1}_{2N \times 2N} \otimes \boldsymbol \sigma \otimes \gamma_0 \psi$, 
where $\mathbf{\Omega}(x)$ is a three-component, vector field. A uniform order parameter field determines a global spin quantization axis along $\mathbf{\Omega}=\mathbf{\Omega}_0$ and the total $U(1)$ spin-current $J^s_{\mu}= \hat{\mathbf{\Omega}}_0 \cdot \bar{\psi} \mathbb{1}_{2N \times 2N} \otimes \boldsymbol \sigma \otimes \gamma_\mu \psi$. In response to the externally applied electro-magnetic and spin gauge fields, the QSH phase supports cross-correlated, charge and spin Hall currents, $J^0_{\alpha}=\frac{\sigma_{xy}^{s}}{2\pi} \;  \epsilon_{\alpha \mu \nu} F^0_{\mu \nu}$, and $J^s_{\alpha}=\frac{\sigma_{xy}^{s}}{2\pi} \;  \epsilon_{\alpha \mu \nu} F^s_{\mu \nu}$
where $\sigma^{s}_{xy} =  2N$ is the quantized spin-Hall conductivity, and the Abelian field strength tensors are $F^i_{\mu \nu}=\partial_\mu A^i_\nu - \partial_\nu A^i_\mu$. 
Similar results for the cross-correlated charge and flavor quantum Hall currents can be obtained for QVH and QMVH mass orders in TBLG,  respectively defined as
$\mathcal{O}_{QVH} = \mathbf{\Omega} (x) \cdot \bar{\psi} \boldsymbol \eta \otimes \rho_0 \otimes \sigma_0  \otimes \gamma_0 \psi$, and 
$\mathcal{O}_{QMVH} = \mathbf{\Omega} (x) \cdot \bar{\psi} \eta_0 \otimes \boldsymbol \rho \otimes \sigma_0  \otimes \gamma_0 \psi$. 
 
The static skyrmion configurations of QSH order parameter are described by 
\begin{equation}
\hat{\mathbf{\Omega}}(\rho, \varphi)=\left [\sin  \theta \cos(n \varphi), \sin  \theta \sin ( n \varphi), \cos \theta \right],\label{eq2}
\end{equation}
where $\varphi= \tan^{-1}(x_2/x_1)$, and $\theta$ is a function of radial variable $\rho= \sqrt{x^2_1 + x^2_2}$, such that $\theta(\rho \to 0) \to 0$ and $\theta(\rho \to \infty) \to \pi$. The winding number $W_{sk}=\frac{1}{4\pi} \; \int d^2x \;  \hat{\mathbf{\Omega}} \cdot (\partial_{\nu} \hat{\mathbf{\Omega}} \times \partial_\lambda  \hat{\mathbf{\Omega}}) \; = n$, with $n=\pm 1, \pm 2,..$. The Belavin-Polyakov solutions of skyrmions,~\cite{Polyakov} which minimize the energy of static NLSMs, correspond to 
$\cos(\theta)=(R^n - \rho^n)/(R^n + \rho^n)$, where $R$ is the size of skyrmion core.

Deep inside the ordered phase, the induced fermion current can be computed by employing the gradient expansion scheme, which is controlled by the amplitude $|\mathbf{\Omega}|$. The spin-singlet, total number current is found to be 
\begin{equation}
J^0_{\mu}= \sigma^s_{xy} J^{sk}_{\mu} = 2N J^{sk}_{\mu}, \label{eq3}
\end{equation} where $J^{sk}_{\mu}= \frac{1}{4\pi} \; \epsilon_{\mu \nu \lambda} \hat{\mathbf{\Omega}} \cdot (\partial_{\nu} \hat{\mathbf{\Omega}} \times \partial_\lambda  \hat{\mathbf{\Omega}})$ is the skyrmion current density. Since the ordered phase does not allow singular tunneling events or hedgehog configurations, both $J^{sk}_{\mu}$ and $J^0_{\mu}$ satisfy the continuity equation. Hence, the induced fermion number $
\langle \bar{\psi} \gamma_0 \psi \rangle =2N \int d^2x J^{sk}_{0}=2N \; W_{sk}$, 
is determined by the quantized, spin Hall conductivity, and $W_{sk}$ acts as the generator of flavor-singlet, $U(1)$ symmetry. 

As the $SU(2N)$ flavor symmetry remains unbroken, the spin-singlet $SU(2N)$ flavor currents $J^{0,l}_{\mu}= \bar{\psi} \hat{\lambda}_l \otimes \sigma_0 \otimes \gamma_\mu \psi$, with $l=1,..,(2N)^2-1$ remain identically conserved. 
When the QSH order is destroyed, the condensation of skyrmions is expected to give rise to a spin-singlet, paired state, breaking global $U(1)$ symmetry. Similarly, the valley- (mini-valley-) singlet paired states can arise from the quantum-disordered QVH (QMVH) phase. \emph{However, all charge $2e^-$ fermion bilinears break $SU(2N)$ flavor symmetry, when $N>1$} (see Appendix~\ref{App1}). \emph{Can the skyrmion condensation determine the pattern of flavor-symmetry-breaking (FSB) by paired states}? To answer this question, we consider the role of hedgehog configurations, which serve as the source and sink of $J^{sk}_{\mu}$.

\underline{\emph{Hedgehogs and fermion zero modes}}:
These tunneling singularities in Euclidean space-time are also classified according to the second homotopy group $\Pi_2(S^2)=\mathbb{Z}$. The topological invariant or hedgehog charge is given by 
$q=\frac{1}{8\pi} \int d^2S_{a} \epsilon_{abc} \; \epsilon_{\alpha \beta \lambda} \; \hat{\Omega}_{\alpha} \partial_b \hat{\Omega}_{\beta} \partial_c \hat{\Omega}_{\lambda}$,
where the integrals are performed over a sphere surrounding the singularity.~\cite{Arafune} The minimal or unit strength ($q=\pm 1$), radial (anti-)hedgehog configurations correspond to $\hat{\Omega}_\mu = \pm \hat{x}_\mu$. More general hedgehogs with charge $q_h = l$ can be described by 
\begin{eqnarray}
\mathbf{\Omega}=\left[F_{1,l} (\rho) \cos(l \varphi), F_{1,l} (\rho) \sin(l \varphi), F_{2,l}(x_0) \right],\label{eq4}
\end{eqnarray}
where $F_{1,l}$ and $F_{2,l}$ are two profile functions with the asymptotic properties: (i) $F_{1,l} (\rho \to 0) \sim \rho^l$, and $F_{1, l} (\rho \to \infty) \sim c_1$, and (ii) $F_{2,l} (x_0 \to 0) \sim x_0$ and $F_{2,l} (x_0 \to \pm \infty) \sim \pm c_2$, with $c_1$ and $c_2$ being two constants.
Rotating these configurations by an angle $\phi$ about a constant unit vector $\hat{\mathbf{n}}$, we can find various topologically equivalent configurations 
$\hat{\mathbf{\Omega}}^\prime=\mathcal{R}(\hat{\mathbf{n}},\phi) \hat{\mathbf{x}}$. 
where $\mathcal{R}(\hat{\mathbf{n}},\phi)$ is a rotation matrix. 

Due to the non-conservation of skyrmion currents, the continuity equation for the total number current is modified to
\begin{eqnarray}
\partial_\mu J^0_{\mu}
=2N \sum_{i} \;\delta^3(x-x_{i}) q_{i} .\label{eq5}
\end{eqnarray}
To satisfy the overall neutrality condition (i.e., $\sum_i q_{i}=0$), we must have an equal number of hedgehogs and anti-hedgehogs. A fully self-consistent treatment of fermion-hedgehog interactions lies beyond the scope of current work. We will only discuss the topological structure of fermion determinant in the presence of fixed hedgehog configurations. 

Let us consider the coupling between hedgehogs of QSH order parameter and one species of four-component Weyl fermions with sub-lattice and spin index. Due to the $SU(2N)$ flavor symmetry, the results for all species of Weyl fermions will be identical.
To facilitate our analysis of fermion-hedgehog scattering and induced pairing, we introduce the Nambu spinor $\Psi^T_N= (\psi^T, \bar{\psi}\gamma_1 \otimes \sigma_2)/\sqrt{2}$ and $\bar{\Psi}_N= (\psi^T \gamma_1 \otimes \sigma_2, \bar{\psi})/\sqrt{2}$, leading to 
\begin{eqnarray}
S=\int d^3x \; \bar{\Psi}_N \mathcal{D}_N \Psi_N=\int d^3x \; \bar{\Psi}_N \left(\begin{array}{cc}
0 & \mathcal{D}^\dagger\\
\mathcal{D} & 0
\end{array}\right) \Psi_N,
\end{eqnarray} where 
$\mathcal{D}=\gamma_\mu \partial_\mu +  \boldsymbol \sigma \cdot \bs{\Omega} $. In this basis, all charge $2 e^-$, fermion bilinears will be block diagonal operators (see Appendix~\ref{App1}). While $\mathcal{D}$ and $\mathcal{D}^\dagger$ are not Hermitian or anti-Hermitian operators, $\mathcal{D}_N$ is an Hermitian operator. Therefore, the fermion fields $\Psi_N$ and $\bar{\Psi}_N$ can be expanded in the eigenbasis of $\mathcal{D_N}$. 
The eigenvalues follow from the squared operator \begin{equation}
\mathcal{D}^2_N=\left(\begin{array}{cc}
\mathcal{D}^\dagger\mathcal{D} & 0\\
0 & \mathcal{D}\mathcal{D}^\dagger
\end{array}\right). \end{equation} 
Since $\{ \mathcal{D}_N, \alpha_3 \} = 0$, where $\alpha_3$ is the diagonal Pauli matrix, operating on the particle-hole index, the non-zero eigenvalues of $\mathcal{D}_N$ come in pairs. 

However, the zero-modes possess definite chirality, corresponding to  $\alpha_3= +1$ (right/annihilation channel) or $-1$ (left/creation channel). 
The four-component, right zero-mode $\phi_{R}$ (column vector) satisfies $\mathcal{D} \phi_R=\mathcal{D}^\dagger \mathcal{D} \phi_{R}=0$ and the four-component, left zero-mode $\phi_{L}$ (row vector) obeys $\mathcal{D}^\dagger \phi_L^\dagger=\mathcal{D} \mathcal{D}^\dagger\phi_L^\dagger=0$. The difference between the number of right zero-modes $n_R$ and the number of left zero-modes $n_L$ is always -$|l|$ for hedgehogs and +$|l|$ for anti-hedgehogs. \emph{After accounting for the $2N$-fold degeneracy in flavor index, we find the total number of topologically protected fermion zero-modes is equal to $\sigma^s_{xy} \times |l| = 2N |l|$.} This result is protected by Callias's index theorem.~\cite{JackiwRebbi1, Callias1, Callias2}

For the unit-strength, radial (anti-)hedgehog configurations, with $\bs{\Omega}_\mu= \pm \; m(x) \; \hat{x}_\mu$, where $m$ is amplitude of the order parameter, the zero-mode eigenfunctions are given by
\begin{equation}
\phi^\ast_{+,L}=\phi_{-,R}^T=(0,1,-i,0)f(|x|),\label{eq7}
\end{equation}
where $f(|x|) \propto \exp[- \int dx^\prime m(x^\prime)]$. For a constant $m$, we obtain $f(|x|)=\sqrt{\frac{|m|^3}{2\pi}}e^{-m |x|}$.~\cite{GoswamiSi3} In this case, one can also determine the full spectra of $\mathcal{D}$ by following Refs.~\onlinecite{JackiwRebbi1, Callias1}.
The existence of zero-modes for the dynamic Weyl operator indicates the fermion determinant (or Pfaffian) vanishes in the presence of isolated hedgehogs. Therefore, the coupling to Weyl fermions suppresses the probability of any isolated hedgehog. In the vicinity of quantum phase transitions, due to the divergent correlation length, we can consider a dilute gas of hedgehogs. Therefore, the overlap between widely separated zero-modes provides a clear idea about the nature of competing orders, arising from the condensation of skyrmions.

 \begin{figure}[t]
\centering
\includegraphics[scale=0.25]{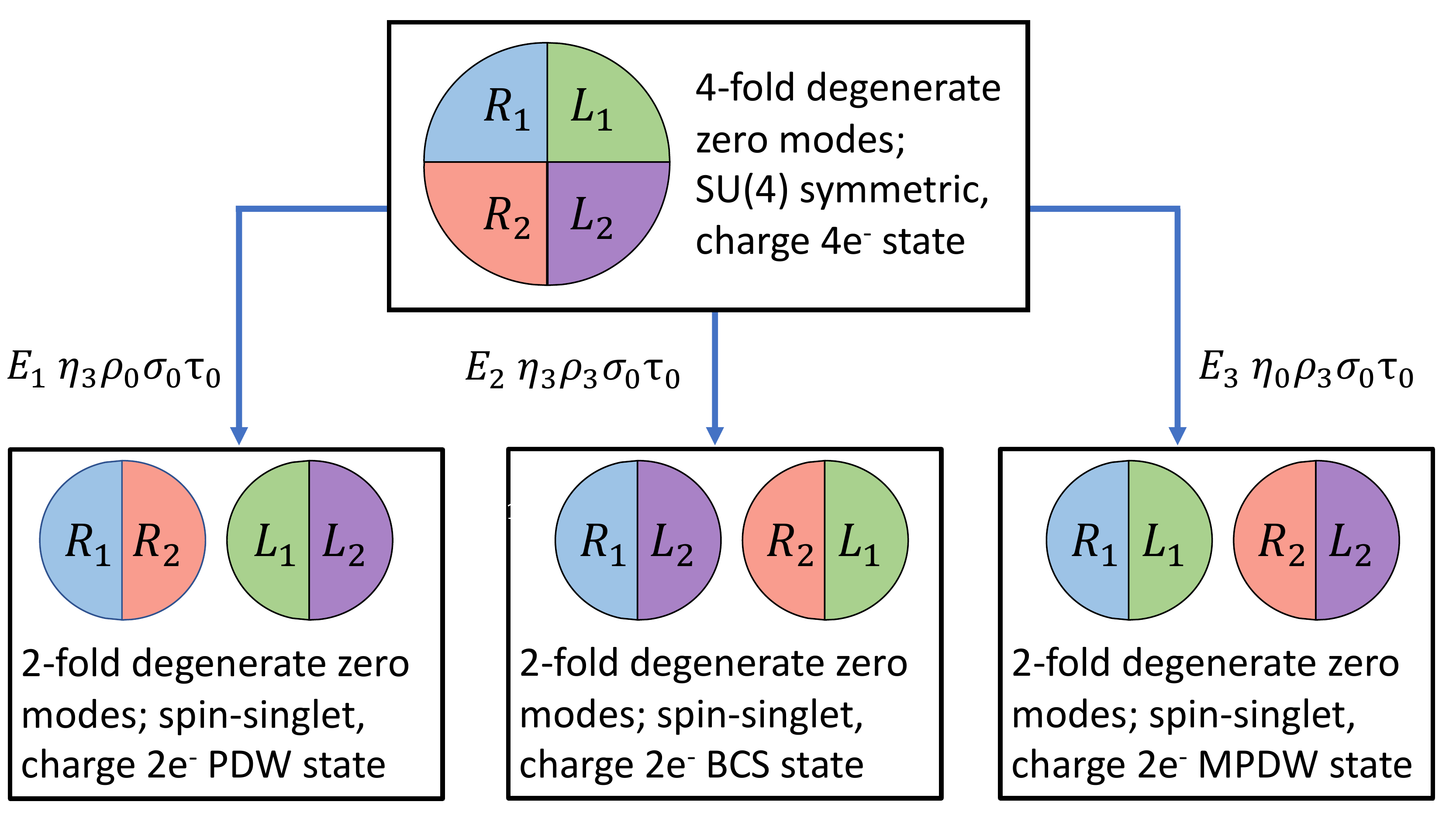}
\caption[]{The degeneracy of fermion zero-modes, localized on hedgehog configurations of quantum spin Hall order parameter and the nature of paired state. Here, $E_1$, $E_2$ and $E_3$ represent three different types of flavor symmetry breaking chemical potentials. } 
\label{Fig1}
\end{figure}

\underline{\emph{'$t$ Hooft vertex and pairing}}: Since a strength $l$ (anti-)hedgehog leads to $2N |l|$ zero-modes in the (annihilation) creation channel, we anticipate a (anti-)hedgehog creation operator will be coupled to $2 N |l|$ number of fermion (annihilation) creation operators. The calculation of such effective TV can be performed by following Refs. \cite{'T Hooft,Shuryak,GoswamiSi3} It is crucial to average over the arbitrary orientations of hedgehogs for a disordered or para-magnetic phase. The eigenfunction of arbitrarily oriented and radial hedgehogs are related by $SU(2)$ rotations $\psi(\hat{\bs{n}}, \phi)=\mathcal{U}^\dagger \psi(\hat{x})$, where $\mathcal{U}=\pm e^{i \phi/2 \hat{\bs{n}} \cdot \boldsymbol \sigma}$, such that $\hat{\mathbf{n}} \cdot \boldsymbol \sigma =\mathcal{U}^\dagger \hat{x} \cdot \boldsymbol \sigma \mathcal{U}$. Since higher-$l$ hedgehogs will have lower probability, we will only consider TV due to $l=\pm 1$ hedgehogs.

After performing the integral over $SU(2)$ group, the two-fermion, TV for MLG with $N=1$ becomes
\begin{equation}
Y=y \int d^3k f^2(k)k^2 \epsilon^{ij}[\psi_i^T(-k)\psi_j(k)+\bar{\psi}(k)\bar{\psi}^T_j(-k)], \label{eqTV1}
\end{equation}
where $y$ is the fugacity of $l = \pm 1$ hedgehogs, $i,j\in\{R,L\}$, and $\psi^T$ and $\bar{\psi}^T$ have $\gamma_1 \otimes\sigma_2$ absorbed by definition. The form factor $f(k)=\frac{4\sqrt{2\pi}m^{5/2}}{(k^2+m^2)^2}$ is the Fourier transform of $f(r)=\sqrt{\frac{|m|^3}{2\pi}}e^{-|m|r}$. The TV describes \emph{frequency-momentum dependent charge $2e^-$ bilinear in the spin-singlet, $s$-wave pairing channel}. Due to the exponentially localized behavior of zero-mode eigenfunctions, $f(k)$ exhibits short-ranged behavior. Since the minimal hedgehogs induce charge $2e^-$ pairing mass, the use of level-$1$ $SO(5)$ WZW theory for describing competing orders is justified.

\indent For TBLG with $N=2$, the calculation of quartic TV due to four degenerate zero-modes of minimal hedgehogs requires some algebraic manipulations. The final result is given by
\begin{align}
\begin{split}
Y=&y\int d^3k_1 \int d^3k_2 f^2(k_1)f^2(k_2)k_1^2k_2^2\\
&\epsilon^{ijkl}[\psi^T_i(-k_1)\psi_j(k_1)\psi^T_k(-k_2)\psi_l(k_2)+h.c.], \label{eq9}
\end{split}
\end{align}
where $i,j\in\{R_1,R_2,L_1,L_2\}$ and $\psi^T$ and $\bar{\psi}^T$ have $\gamma_1 \otimes\sigma_2$ absorbed by definition. \emph{The quartic, TV describes spin-singlet, composite, charge $4e^-$ paired state that preserves $SU(4)$ symmetry in the combined valley and mini-valley space (enforced by $\epsilon^{ijkl}$)}.   Obviously such a phase cannot be described by fermion bilinears. Similar conclusions can be drawn for the quantum-disordered QVH and QMVH phases. \emph{In summary, hedgehogs only break $U(1)$ symmetry and induce pairing, without breaking flavor symmetry.}    

\underline{\emph{Effects of flavor-symmetry-breaking}}: In order to realize charge $2e^-$ states for $N \geq 2$, the $SU(2N)$ symmetry must be broken through additional mechanism. In contrast to the long-range tail of Coulomb interactions, the generic short-range interactions do not respect flavor symmetry, inherited from the valley/mini-valley degrees of freedom. Thus, depending on many non-universal details of short-range interactions, different types of charge $2e^-$ states can emerge out of charge $4e^-$ paired states.  

The FSB by high-temperature orders can also be instrumental in selecting charge $2e^-$ states. For example, we can consider flavor dependent chemical potentials, which shift various Weyl points to different reference energies. Such chemical potentials couple to the flavor-density operators and reduce the strength of $\sigma^s_{xy}$ from being $2$ to $1$. For TBLG, in addition to the conventional chemical potential ($E_0$) for the total number operator, we can consider three types of spin-singlet, chemical potentials ($E_j$ with $j=1,2,3$) in the valley and mini-valley space. When $E_0 = \pm E_j$, the degeneracy of zero-modes is reduced to $2$. As illustrated in Fig.~\ref{Fig1}, depending on the explicit nature of FSB, charge $2e^-$ BCS (zero center of mass momentum), pair-density-waves/PDW (large center of mass momentum or short wavelength modulations), and Moi\'re pair-density-waves/MPDW (small center of mass momentum or long wavelength modulations) states can be realized by destroying QSH order. Very similar considerations for QVH and QMVH orders show the possibilities of realizing spin-singlet and spin-triplet, charge $2e^-$ states in BCS and pair-density-wave channels (see Appendix~\ref{App1}).

\underline{\emph{Outlook}}: There are growing experimental and theoretical evidence in favor of QFH order at various integer filling fractions.~\cite{Cao1,Taniguchi,Efetov1,Efetov2,Saito,Kerelsky,Choi,Jiang,Xie,Wong,Zondiner,LiuWang,Cao2,ZhangPo,KangVafek1,KangVafek3,Liao}  Therefore, the condensation of skyrmion textures of QFH insulators can provide valuable insights into the nature of proximate superconducting states, which are not easily accessible from weak or strong coupling analysis of microscopic models. The experiments also suggest the presence of FSB chemical potentials, which can cause revival of Weyl points, in the vicinity of Fermi level.~\cite{Wong,Zondiner} More concrete evidence for the precise nature of QFH order and FSB chemical potentials are required for making definitive predictions for pairing symmetry in TBLG. Our work suggests exciting possibilities of realizing exotic charge $4e^-$ states around $\nu=0$ and pair-density-waves for all filling fractions through topological pairing mechanism.

\acknowledgements{
This work was supported by the National Science Foundation MRSEC program (DMR-1720139) at the Materials Research Center of Northwestern University.
}

\appendix

\section{Pairing bilinears}\label{App1}

In this appendix, we outline FSB fermion bilinears in particle-particle channels. The relationship with appropriate $SO(5)$ and $SO(9)$ NLSMs of competing orders will be identified.
For the general case of $4N$-flavors of Weyl fermions, all momentum-independent, charge $2e^-$ pairing terms can be described as $\phi_{l,1} \Psi^\dagger \alpha_1 \hat{\Lambda}^\prime_l \Psi$ and $\phi_{l,2} \Psi^\dagger \alpha_2 \hat{\Lambda}^\prime_l \Psi$, where $\Psi^\dagger=(\psi^\dagger, \psi^T)/\sqrt{2}$ is the $16N$-component Nambu spinor, $\hat{\Lambda}^\prime_l$ are the $4N(8N-1)$ number of $8N \times 8N$, imaginary, $SU(8N)$ Gell-Mann matrices, and the Pauli matrices $\alpha_j$'s operate on the particle-hole index. Notice that the number of pairing bilinears $4N(8N-1)$ exactly equals the number of generators for $SO(8N)$ group. Out of these bilinears, only $\Psi^\dagger \alpha_j \otimes \hat{M} \otimes \tau_2 \Psi$, with symmetric $4N \times 4N$ matrix $\hat{M}$ can anti-commute with the Hamiltonian of free fermions and act as superconducting mass terms. There are $2N(4N+1)$ independent, charge $2e^-$, mass terms, which is equal to the number of generators for $USp(4N)$ group. By absorbing the imaginary Pauli matrices ($\eta_2$, $\rho_2$ etc.) from all $SU(2)$ sectors on the lower component of Nambu spinor, the pairing mass terms can be organized in a more convenient singlet and triplet forms.

\subsection{Spinful MLG} For MLG with $N=1$, we will absorb $\eta_2\otimes\sigma_2\otimes\tau_2$ into the lower component of Nambu spinor: $\Psi^T_1=(\psi_1,\tilde{\psi}^*_1)^T/\sqrt{2}=(\psi_1,\eta_2\otimes\sigma_2\otimes\tau_2\bar{\psi}^T_1)^T/\sqrt{2}$, and define the barred Nambu spinor with an additional $\alpha_1$ absorbed: $\bar{\Psi}_1=\alpha_1(\bar{\psi}_1,\tilde{\psi}^T_1)/\sqrt{2}=(\psi^T_1\eta_2\otimes\sigma_2\otimes\tau_2,\bar{\psi}_1)/\sqrt{2}$. In this basis, the QSH and QVH mass terms become 
\begin{align}
\mathcal{O}_{QSH}&=\mathbf{\Omega}_1 \cdot \bar{\Psi} \alpha_1\alpha_3\otimes\eta_0\otimes\boldsymbol \sigma\otimes\tau_0 \Psi, \\
\mathcal{O}_{QVH}&= \mathbf{\Omega}_2 \cdot \bar{\Psi} \alpha_1\alpha_3\otimes \boldsymbol \eta \otimes\sigma_0\otimes\tau_0 \Psi, 
\end{align}
while the Hamiltonian operator of free fermion acquires the form $H_f= - i \sum_{j=1}^{2} \; \bar{\Psi} \alpha_1\alpha_3 \otimes \mathbb{1}_{4 \times 4} \otimes  \gamma_j \partial_j \Psi$. The ten possible pairing masses can be rewritten as 
\begin{align}
\mathcal{O}_{s,j}&=\phi_{s,j} \bar{\Psi} \alpha_1\alpha_j \otimes \eta_0 \otimes \sigma_0 \otimes \tau_0 \Psi, \\
\mathcal{O}^{mn}_{t,j}&=\phi_{t,j}^{mn} \bar{\Psi} \alpha_1\alpha_j \otimes \eta_m \otimes \sigma_n \otimes \tau_0 \Psi, 
\end{align}
with $m$ and $n$ being equal to $1,2,3$. Only the spin and valley singlet, $s$-wave, pairing mass $\mathcal{O}_{s,j}$ can anti-commute with $H_f$, $\mathcal{O}_{QSH}$ and $\mathcal{O}_{QVH}$. Hence, the condensation of skyrmion textures of both QSH and QVH order parameters in the para-magnetic phase can lead to a unique, charge $2e^-$, spin- and valley- singlet, $s$-wave pairing mass. Consequently, we can construct two quintuples 
\begin{eqnarray}
\mathbf{N}_{1}&=&(\Omega_{1,1}, \Omega_{1,2}, \Omega_{1,3}, \phi_{s,1}, \phi_{s,2}), \\
\mathbf{N}_{2}&=&(\Omega_{2,1}, \Omega_{2,2}, \Omega_{2,3}, \phi_{s,1}, \phi_{s,2}), 
 \end{eqnarray} 
for describing competition between particle-hole and particle-particle channels. Simple matrix algebra shows the resulting $SO(5)$ NLSMs support level-$1$ WZW term. \emph{This is in agreement with the TV being a charge $2e^-$ bilinear} [see Eq.~(\ref{eqTV1}) ].  Since the spin and valley triplet pairing mass terms $\mathcal{O}^{mn}_{t,j}$ do not anti-commute with $\mathcal{O}_{QSH}$ or $\mathcal{O}_{QVH}$, they are not favored as emergent, pairing orders.

\begin{table}[t]
	\def\arraystretch{1.5}
	\begin{tabular}{|c|c|c|c|}
		\hline
		QFH & QSH & QVH & QMVH\\
		\hline
		PDW & $\mathcal{O}_{vt,j}^{100}$, $\mathcal{O}_{vt,j}^{200}$ & N/A & $\mathcal{O}_{vt,j}^{100}$, $\mathcal{O}_{vt,j}^{200}$\\
		\hline
		MPDW & $\mathcal{O}_{mt,j}^{010}$, $\mathcal{O}_{mt,j}^{020}$ & $\mathcal{O}_{mt,j}^{010}$, $\mathcal{O}_{mt,j}^{020}$ & N/A\\
		\hline
		BCS & $\mathcal{O}_{vt,j}^{300}$, $\mathcal{O}_{mt,j}^{030}$ & $\mathcal{O}_{mt,j}^{030}$, $\mathcal{O}_{st,j}^{00l}$ & $\mathcal{O}_{vt,j}^{300}$, $\mathcal{O}_{st,j}^{00l}$\\
		\hline
	\end{tabular}
\caption{Classification of pairing mass terms for each QFH order, with j=1,2 and l=1,2,3.} \label{tab1}
\end{table}

\subsection{Spinful TBLG} For TBLG with $N=2$, we absorb $\eta_2\otimes\rho_2\otimes\sigma_2\otimes\tau_2$ into the lower component of Nambu spinor: $\Psi^T_2=(\psi_2,\tilde{\psi}^*_2)^T/\sqrt{2}=(\psi_2,\eta_2\otimes\rho_2\otimes\sigma_2\otimes\tau_2\bar{\psi}^T_2)^T/\sqrt{2}$, and define the barred Nambu spinor with an additional $\alpha_1$ absorbed: $\bar{\Psi}_2=\alpha_1(\bar{\psi}_2,\tilde{\psi}^T_2)/\sqrt{2}=(\psi^T_2\eta_2\otimes\rho_2\otimes\sigma_2\otimes\tau_2,\bar{\psi}_2)/\sqrt{2}$. In this basis, the QSH, QVH, QMVH mass terms become 
\begin{align}
\mathcal{O}_{QSH}&=\mathbf{\Omega}_1 \cdot \bar{\Psi} \alpha_1\alpha_3\otimes\eta_0\otimes\rho_0\otimes\tau_0\otimes\boldsymbol \sigma \Psi, \\
\mathcal{O}_{QVH}&= \mathbf{\Omega}_2 \cdot \bar{\Psi} \alpha_1\alpha_3\otimes \boldsymbol \eta \otimes\rho_0\otimes\tau_0\otimes\sigma_0 \Psi, \\
\mathcal{O}_{QMVH}&= \mathbf{\Omega}_3 \cdot \bar{\Psi} \alpha_1\alpha_3\otimes  \eta_0 \otimes\boldsymbol\rho\otimes\tau_0\otimes\sigma_0 \Psi, 
\end{align}
while the Hamiltonian operator of free fermion acquires the form $H_f= - i \sum_{j=1}^{2} \; \bar{\Psi} \alpha_1\alpha_3 \otimes \mathbb{1}_{8 \times 8} \otimes  \gamma_j \partial_j \Psi$. The pairing mass terms can be grouped into four categories: 
\begin{align}
 \mathcal{O}_{st,j}^{00l}&=\phi_{st,j}^{00l} \bar{\Psi} \alpha_1\alpha_j  \otimes \eta_0 \otimes \rho_0 \otimes \sigma_l \otimes \tau_0 \Psi, \\
 \mathcal{O}_{vt,j}^{l00}&=\phi_{vt,j}^{l00} \bar{\Psi} \alpha_1\alpha_j  \otimes \eta_l \otimes \rho_0 \otimes \sigma_0 \otimes \tau_0 \Psi, \\
 \mathcal{O}_{mt,j}^{0l0}&=\phi_{mt,j}^{0l0} \bar{\Psi} \alpha_1\alpha_j  \otimes \eta_0 \otimes \rho_l \otimes \sigma_0 \otimes \tau_0 \Psi, \\
 \mathcal{O}_{at,j}^{lmn}&=\phi_{at,j}^{lmn} \bar{\Psi} \alpha_1\alpha_j  \otimes \eta_l \otimes \rho_m \otimes \sigma_n \otimes\tau_0 \Psi,
 \end{align} 
where $l$, $m$, and $n$ can take values $1,2,3$. They respectively describe pairing bilinears, which are spin-triplet, valley-triplet, mini-valley-triplet, and triplets in all flavor channels. The physical significance of such pairing mass terms are described in Table~\ref{tab1}.

The QSH mass for $N=2$ model anti-commutes with charge $2 e^-$ bilinears $\mathcal{O}_{vt}^{l00}$ and $\mathcal{O}_{mt}^{0l0}$. There are six possible ways to form \emph{quintuples} 
\begin{eqnarray}
\mathbf{N}^{l}_{1}&=&(\Omega_{1,1}, \Omega_{1,2}, \Omega_{1,3}, \phi_{vt,1}^{l00}, \phi_{vt,2}^{l00}), \\
 \mathbf{N}^{l}_{2}&=&(\Omega_{1,1}, \Omega_{1,2}, \Omega_{1,3}, \phi_{mt,1}^{0l0}, \phi_{mt,2}^{0l0}),
 \end{eqnarray} 
with $l=1,2,3$, and \emph{all six types of $SO(5)$ NLSMs will support level-$2$ WZW term}. Due to the presence of multiple pairing mass terms, the skyrmion condensation cannot select any specific charge $2e^-$ pairing channel, without any additional mechanism of breaking $SU(4)$ flavor symmetry. 

It is possible to combine all competing mass terms into a \emph{nonuple} 
\begin{equation}
\mathbf{L}=(\Omega_1, \Omega_2, \Omega_3, \phi_{vt,1}^{100}, \phi_{vt,1}^{200}, \phi_{vt,1}^{300}, \phi_{mt,2}^{010}, \phi_{mt,2}^{020}, \phi_{mt,2}^{030}),
\end{equation} 
and write down $SO(9)$ NLSM for $\hat{\mathbf{L}}$. The $SO(9)$ NLSM treats all mutually anti-commuting mass orders on an equal footing. The resulting coset space $SO(9)/SO(8)$ is closely tied to the octonion gauge theories and details of such exotic aspects will be provided elsewhere. Within the $SO(9)$ model, the pairing mass terms are embedded as a \emph{sextuple}. The appearance of $SO(6)$ is a natural consequence of $SU(4)$ being the double cover of $SO(6)$. Similar results can be obtained for the combinations $(\mathcal{O}_{QVH}, \mathcal{O}_{st,j}^{00l}, \mathcal{O}_{mt,j}^{0l0})$ and $(\mathcal{O}_{QMVH}, \mathcal{O}_{vt,j}^{l00}, \mathcal{O}_{st,j}^{00l})$. This type of \emph{triality} is a consequence of the underlying $SU(8)$ flavor symmetry. \\

\end{document}